\documentstyle[12pt,epsfig]{article}
\def\beq{\begin{equation}}
\def\eeq{\end{equation}}
\def\bea{\begin{eqnarray}}
\def\eea{\end{eqnarray}}
\def\non{\nonumber}
\def\bib{\bibitem}

\def\v{\vert}

\def\r{\rangle}

\begin{document}

\begin{center}
{\large \bf \sf Novel multi-band quantum soliton states for a derivative \\
nonlinear Schr\"odinger model}

\vspace{1.3cm}

{\sf B. Basu-Mallick$^1$\footnote{e-mail address: biru@theory.saha.ernet.in},
Tanaya Bhattacharyya$^1$\footnote{e-mail address: tanaya@theory.saha.ernet.in}
and Diptiman Sen$^2$\footnote{e-mail address: diptiman@cts.iisc.ernet.in}}

\bigskip

{\em $^1$Theory Group, Saha Institute of Nuclear Physics, \\
1/AF Bidhan Nagar, Kolkata 700 064, India} 

\bigskip

{\em $^2$Centre for Theoretical Studies, Indian Institute of Science, \\
Bangalore 560012, India}
\end{center}

\bigskip
\bigskip

\noindent {\bf Abstract}

We show that localized $N$-body soliton states exist for a quantum integrable 
derivative nonlinear Schr\"odinger model for several non-overlapping
ranges (called bands) of the coupling constant $\eta$. The number of such 
distinct bands is given by Euler's $\phi$-function which appears in the 
context of number theory. The ranges of $\eta$ within each band can also be 
determined completely using concepts from number theory such as Farey sequences
and continued fractions. We observe that $N$-body soliton states appearing 
within each band can have both positive and negative momentum. Moreover, for 
all bands lying in the region $\eta >0$, soliton states with positive momentum 
have positive binding energy (called bound states), while the states with 
negative momentum have negative binding energy (anti-bound states).

\bigskip



\newpage

\noindent \section{Introduction}
\renewcommand{\theequation}{1.{\arabic{equation}}}
\setcounter{equation}{0}

\medskip

Soliton states in integrable quantum field theory models in 1+1 dimensions have
been studied extensively for many years 
\cite{thac,fadd,skyl1,skyl2,bhat,shni,kund,basu1,basu2,basu3}. The quantum 
soliton states are usually constructed by using either the coordinate Bethe 
ansatz or the algebraic Bethe ansatz. For an integrable nonrelativistic 
Hamiltonian, the coordinate Bethe ansatz can yield the exact eigenfunctions in
the coordinate representation. If such an eigenfunction decays sufficiently 
fast when any of the particle coordinates tends towards infinity (keeping the 
center of mass coordinate fixed), we call such a localized square-integrable 
eigenfunction a quantum soliton state. It is also possible to construct quantum
soliton states using the algebraic Bethe ansatz, by choosing 
appropriate distributions of the spectral parameter in the complex plane 
\cite{fadd,skyl1,skyl2}. The stability of quantum soliton states, in the 
presence of small external perturbations, can be determined by calculating 
their binding energy. It is usually found that localized quantum soliton states
of various integrable models, including the well known nonlinear Schr\"odinger
model (NLS) and the sine-Gordon model, have positive binding energy 
\cite{thac,fadd,skyl1,skyl2,bhat}. 

In this paper, we will study the quantum soliton 
states of an integrable derivative nonlinear Schr\"odinger 
(DNLS) model \cite{shni,kund,basu1,basu2}. Classical and quantum versions of
the DNLS model have found applications in different areas of physics like 
circularly polarized nonlinear Alfven waves in a plasma \cite{wada,clar},
quantum properties of solitons in optical fibers \cite{rose}, and 
some chiral Luttinger liquids which are obtained by dimensional
reduction of a Chern-Simons model defined in two dimensions \cite{agli,bene}.
It is known that the classical DNLS model can have 
solitons with momenta in only one direction, where the direction
depends on the sign of the coupling constant; namely, the ratio of the momentum
to the coupling constant is positive \cite{chen,kaup,min}. Here we 
want to investigate whether this chirality property of the classical solitons 
is preserved at the quantum level. The Lagrangian and Hamiltonian of the 
quantum DNLS model in its second quantized form are given by \cite{shni}
\bea
L &=& i \int_{-\infty}^{\infty} dx ~\psi^\dagger \psi_t ~-~ H ~, \non \\
H &=& \int_{-\infty}^{+\infty} dx ~\Big[ ~\hbar \psi_x^{\dagger} \psi_x +
i\eta \{{\psi^\dagger}^2 \psi \psi_x - \psi^\dagger_x {\psi^\dagger} \psi^2 
\} ~\Big] ~,
\label{a1}
\eea
where the subscripts $t$ and $x$ denote partial derivatives with respect to 
time and space respectively, $\eta ~ (\neq 0)$ is the real coupling constant, 
and we have set the particle mass $m = 1/2$. The field operators $ \psi(x,t),~
\psi^\dagger(x,t)$ obey the equal time commutation relations,
$[\psi(x,t), \psi(y,t)] = [\psi^\dagger(x,t), \psi^\dagger(y,t)] = 0$, and 
$[\psi(x,t), \psi^\dagger(y,t)]$ $= \hbar \delta (x-y)$. The quantum soliton 
states of this DNLS model have been constructed through both the algebraic 
Bethe ansatz \cite{kund,basu1,basu2} and the coordinate Bethe ansatz 
\cite{shni}. By applying the coordinate Bethe ansatz, it was found that 
quantum $N$-body soliton (henceforth called $N$-soliton) states exist for 
this DNLS model provided that $\eta$ lies in the range
$0 < \v \eta \v < \tan \left( \frac{\pi}{N} \right)$.
Moreover it was observed that, similar to the classical case,
such $N$-soliton states can have only positive values of $P/\eta$, where $P$ is
the momentum \cite{shni}. However, it was found recently that soliton states 
can exist even with $P/\eta <0$ provided that $\tan \left( \frac{\pi}{N} 
\right) < \v \eta \v < \tan \left( \frac{\pi}{N-1} \right)$ \cite{basu3}. These
soliton states have the surprising feature that their binding energy is 
negative. This naturally leads to the question of whether there are additional
ranges of values of $\eta$ for which there are $N$-solitons in the quantum DNLS
model, and if so, what the values of momentum and binding energy of those
solitons are.

In this paper, we re-investigate the ranges of values of $\eta$ for which 
localized quantum $N$-soliton states exist in the DNLS model. In Sec. 2, we 
apply the coordinate Bethe ansatz and find the condition which the Bethe 
momenta have to satisfy in order that a quantum $N$-soliton state should 
exist. In Sec. 3, we find that there are certain nonoverlapping
ranges of $\eta$, called bands, in which solitons can exist. The $N$-solitons 
of DNLS model found earlier [6,10] are contained only within the lowest 
band. Thus the existence of solitons in higher bands is a novel feature of 
DNLS model which is revealed through our present investigation. We also apply 
the idea of Farey sequences in number theory to completely determine the 
ranges of all bands for which $N$-soliton states exist for a given value of 
$N$. In Sec. 4, we show that the solitons appearing within 
each band can have both positive and negative
values of $P/\eta$; these have positive and negative binding energies 
respectively, and we call them bound and anti-bound states. In Sec. 5, we will
use another concept from number theory, that of continued fractions, to give an
explicit expression for the end points of the bands. We will also address the 
inverse problem of finding the values of $N$ for which $N$-soliton states 
exist for a given value of $\eta$. We make some concluding remarks in Sec. 6.

\vspace{1cm}

\noindent \section{Conditions for quantum $N$-soliton states in DNLS model}
\renewcommand{\theequation}{2.{\arabic{equation}}}
\setcounter{equation}{0}

\medskip

To apply the coordinate Bethe ansatz, we separate the full bosonic Fock space 
associated with the Hamiltonian (\ref{a1}) into disjoint $N$-particle subspaces
$\vert S_N \rangle $. We want to solve the eigenvalue equation $H \v S_N \r = E
\v S_N \r $. The coordinate representation of this equation is given by
\bea
H_N ~\tau_N( x_1, x_2, \cdots , x_N ) ~=~ E ~\tau_N( x_1, x_2, \cdots , x_N )~,
\label{b1}
\eea
where the $N$-particle symmetric wave function $ \tau_N( x_1, x_2,\cdots , 
x_N)$ is defined as 
\bea
\tau_N( x_1, x_2, \cdots , x_N ) ~=~ \frac{1}{\sqrt{n!}} ~\langle 0 \vert 
\psi(x_1)\cdots \psi(x_N)\vert S_N \rangle ~,
\label{b2}
\eea 
and $H_N$, the projection of the second-quantized Hamiltonian $H$ 
(\ref{a1}) on to the $N$ particle sector, is given by
\beq
H_N ~=~ -\hbar^2 ~\sum_{j=1}^N ~\frac{\partial^2}{\partial x_j^2} ~+~ 2i 
\hbar^2 \eta ~\sum_{l<m} ~\delta(x_l - x_m )~ \Big( \frac{\partial}{\partial 
x_l} + \frac{\partial} {\partial x_m} \Big).
\label{b3}
\eeq
It is evident that $H_N$ commutes with 
the total momentum operator in the $N$-particle sector, which is defined as
\beq
P_N ~=~ -i\hbar ~\sum_{j=1}^N ~\frac{\partial}{\partial x_j} ~. 
\label{b4}
\eeq

Note that the Hamiltonian and momentum enjoy the scaling property $H_N 
\rightarrow \lambda^2 H_N$ and $P_N \rightarrow \lambda P_N$ if all the 
coordinates $x_i \rightarrow x_i /\lambda$. Given any one eigenfunction of 
$H_N$ and $P_N$, therefore, we can find a one-parameter family of 
eigenfunctions by scaling all the $x_i$. We also observe that $H_N$ remains 
invariant while $P_N$ changes sign if we change the sign of $\eta$ and 
transform all the $x_i \rightarrow - x_i$ at the same time; let us call this 
the parity transformation. Hence it is sufficient to study the problem for
one particular sign of $\eta$, say, $\eta >0$. The eigenfunctions for $\eta < 0$ 
can then by obtained by changing $x_i \rightarrow -x_i$;
this leaves the energy invariant but reverses the momentum.

Next, we divide the coordinate space $R^N \equiv \{ x_1, x_2, \cdots x_N \} $ 
into various $N$-dimensional sectors defined through inequalities like 
$x_{\omega(1)}< x_{\omega(2)}< \cdots < x_{\omega(N)}$, where $\omega(1), 
\omega(2), $ $\cdots , \omega(N)$ represents a permutation of the integers
$1,2, \cdots ,N$. Within each such sector, the interaction part of the 
Hamiltonian (\ref{b3}) is zero, and the resulting eigenfunction is just a 
superposition of the free particle wave functions. The coefficients associated 
with these free particle wave functions can be obtained from the 
interaction part of the Hamiltonian (\ref{b3}), which is nontrivial 
at the boundary of two adjacent sectors. It is known that all such necessary 
coefficients for the Bethe ansatz solution of a $N$-particle 
system can be obtained by solving the corresponding two-particle
problem \cite{gutk}. Let us therefore construct the eigenfunctions 
of the Hamiltonian (\ref{b3}) for the two-particle case, without imposing any
symmetry property on $\tau_2(x_1,x_2)$ under the exchange of the particle 
coordinates. For the region $x_1<x_2$, we may take the eigenfunction to be
\bea
\tau_2 (x_1, x_2) ~=~ \exp ~\{ i(k_1 x_1 + k_2 x_2) \} ~, 
\label{b5}
\eea
where $k_1$ and $k_2$ are two distinct wave numbers. 
Using Eq. (\ref{b1}) for $N=2$, we find that this two-particle 
wave function takes the following form in the region $x_1>x_2 ~$:
\bea
\tau_2(x_1,x_2) ~=~ A(k_1,k_2)\exp ~\{ i(k_1 x_1 + k_2 x_2) \} + B(k_1,k_2)
\exp ~\{ i(k_2 x_1 + k_1 x_2) \} ~,
\label{b6}
\eea
where the `matching coefficients' $A(k_1,k_2)$ and $B(k_1,k_2)$ are given by
\beq
A(k_1,k_2) ~=~ \frac{k_1 - k_2 + i \eta (k_1+ k_2)}{k_1 - k_2} ~, ~~~~
B(k_1,k_2) ~=~ 1- A(k_1 , k_2) ~. 
\label{b7}
\eeq
By using these matching coefficients, we can construct completely symmetric
$N$-particle eigenfunctions for the Hamiltonian (\ref{b3}). In the region
$x_1< x_2 < \cdots < x_N$, these eigenfunctions are given by \cite{shni,gutk}
\beq
\tau_N (x_1, x_2 , \cdots , x_N) ~=~ \sum_\omega \left (\prod_{l<m}
\frac{A(k_{\omega(m)},k_{\omega(l)})}{A(k_m,k_l)}\right) \rho_{\omega(1), 
\omega(2), \cdots , \omega(N)} (x_1, x_2, \cdots , x_N) ~,
\label{b8}
\eeq
where 
\beq
\rho_{\omega(1), \omega(2), \cdots , \omega(N)} (x_1, x_2, \cdots , x_N) ~=~
\exp ~\{ i (k_{\omega(1)}x_1 + \cdots + k_{\omega(N)} x_N ) \} ~.
\label{b9}
\eeq
In the expression (\ref{b8}), the $k_n$'s are all distinct wave numbers, 
and $\sum_{\omega}$ implies summing over all
permutations of the integers $(1,2,....N)$. The eigenvalues of 
the momentum (\ref{b4}) and Hamiltonian (\ref{b3}) operators, corresponding 
to the eigenfunctions $\tau_N(x_1, x_2, \cdots , x_N)$, are given by
\bea
&&~~~~~~~~P_N ~\tau_N(x_1, x_2, \cdots , x_N) ~=~ \hbar 
\Big(\sum_{j=1}^N k_j \Big) ~\tau_N(x_1, x_2, \cdots , x_N) ~, \non
~~~~~~~~~~~~~~~~~~~~ (2.10a) \\
&&~~~~~~~~H_N ~\tau_N(x_1, x_2, \cdots , x_N) ~=~ 
\hbar^2 \Big(\sum_{j=1}^N k_j^2 \Big) ~\tau_N(x_1, x_2, \cdots , x_N) ~. 
\non ~~~~~~~~~~~~~~~~~~~ (2.10b)
\eea
\addtocounter{equation}{1}

The wave function in (\ref{b8}) will represent a localized soliton state if 
it decays when any of the relative coordinates measuring the distance between
a pair of particles tends towards infinity. To obtain the condition for the 
existence of such a localized soliton state, let us consider the following
wave function in the region $x_1<x_2<\cdots <x_N$ :
\bea
\rho_{1,2, \cdots ,N} (~ x_1,x_2, \cdots , x_N ~) ~=~ \exp ~(i\sum_{j=1}^N 
k_j x_j) ~.
\label{b11}
\eea
As before, the momentum eigenvalue corresponding to this wave function is 
given by $\hbar\sum_{j=1}^N k_j$. Since this must be a real quantity, we 
obtain the condition
\beq
\sum_{j=1}^N q_j ~=~ 0 ~,
\label{b12}
\eeq
where $q_j$ denotes the imaginary part of $k_j$. The probability density 
corresponding to the wave function $\rho_{1,2, \cdots ,N} (~ x_1,x_2, \cdots ,
x_N ~)$ in (\ref{b11}) can be expressed as 
\bea
{|\rho_{1,2,\cdots ,N} (~ x_1,x_2, \cdots , x_N ~)|}^2 ~=~ \exp ~\Big\{ ~
2 \sum_{r=1}^{N-1} \Big(\sum_{j=1}^r q_j\Big) ~y_r ~\Big \} ~,
\label{b13}
\eea
where the $y_r$'s are the $N-1$ relative coordinates: $y_r \equiv 
x_{r+1} - x_r$, and we have used Eq. (\ref{b12}). It is evident that the 
probability density in (\ref{b13}) decays exponentially in the limit 
$y_r \rightarrow \infty$ for one or more values of $r$, provided that all 
the following conditions are satisfied:
\beq
q_1< 0 ~, ~~~~q_1+q_2 < 0 ~, ~~\cdots\cdots ~~, ~\sum_{j=1}^{N-1} ~q_j < 0 ~. 
\label{b14}
\eeq
Note that the wave function (\ref{b11}) is obtained by taking $\omega$
as the identity permutation in (\ref{b9}). However, the
full wave function (\ref{b8}) also contains terms like (\ref{b9}) with 
$\omega$ representing all possible nontrivial permutations. 
The conditions which ensure the decay of such a term 
with a nontrivial permutation will, in general, contradict the 
conditions (\ref{b14}). Consequently, in order to have an overall decaying 
wave function (\ref{b8}), the coefficients of all terms 
$\rho_{\omega(1), \omega(2), \cdots , \omega(N)}
(x_1, x_2, \cdots , x_N)$ with nontrivial permutations 
must be required to vanish. This leads to a set of relations like
\beq
A( k_{1}, k_{2} ) ~=~ 0, ~~A( k_{2}, k_{3} ) ~=~ 0, ~\cdots\cdots ~,~~ 
A( k_{N-1}, k_{N}) ~=~ 0 ~.
\label{b15}
\eeq

Thus the simultaneous validity of the conditions
(\ref{b12}), (\ref{b14}) and (\ref{b15}) ensures that the full wave function
$\tau_N(x_1, x_2, \cdots , x_N)$ (\ref{b8}) represents a localized
soliton state. Using the conditions (\ref{b12}) and (\ref{b15}), we 
obtain an expression for the complex $k_n$'s in the form
\beq
k_n ~=~ \chi ~e^{-i(N+1-2n)\phi} ~,
\label{b16}
\eeq
where $\chi$ is a real parameter, and $\phi$ is related to the coupling 
constant as 
\beq
\phi ~=~ \tan^{-1} (\eta ) ~.
\label{b17}
\eeq
To obtain an unique value of $\phi$ from the above equation, we restrict it
to the fundamental region $-\frac{\pi}{2} < \phi (\neq 0) < \frac{\pi}{2}$. 
[Note that $\eta$ and $\phi$ have the same sign.
Due to the parity symmetry mentioned above, we can restrict our 
attention to the range $0 < \phi < \frac{\pi}{2}$].

In this context, it may be mentioned that a relation equivalent to (\ref{b16})
can also be obtained through the method of the algebraic Bethe ansatz, when 
quantum soliton states of DNLS model are considered \cite{kund,basu1,basu2}.

Now, let us verify whether the $k_n$'s in (\ref{b16}) satisfy the conditions 
(\ref{b14}). Summing over the imaginary parts of these $k_n$'s, we can express 
the conditions (\ref{b14}) in the form 
\beq
\chi ~\frac{\sin (l \phi)}{\sin \phi} ~\sin [(N-l) \phi] ~>~ 0 \quad {\rm for}
\quad l ~=~ 1, ~2, ~\cdots ~, N-1 ~. 
\label{b18}
\eeq
Thus, for some given values of $\phi$, $N$ and $\chi$, a soliton state 
will exist when all the above inequalities are simultaneously satisfied.
By using Eqs. (\ref{b16}) and (2.10), we obtain the momentum eigenvalue 
of such soliton state to be
\beq
P ~=~ \hbar \chi ~\frac{\sin (N\phi)}{\sin \phi} ~,
\label{b19}
\eeq
and the energy to be
\beq
E ~=~ \frac{\hbar^2 \chi^2 \sin(2N \phi)}{\sin(2\phi)} ~.
\label{b20}
\eeq
The next section of our paper will be devoted to finding the ranges of values 
of $\phi$ where all the inequalities (\ref {b18})
are simultaneously satisfied for a given value of the particle number $N$.

\vspace{1cm}

\noindent \section{Finding the values of $\phi$ where $N$-soliton states exist}
\renewcommand{\theequation}{3.{\arabic{equation}}}
\setcounter{equation}{0}

\medskip

In this section, we will study the values of $\phi$ where $N$-soliton states 
exist for different values of $N$. For the simplest case $N=2$, 
the condition (\ref{b18}) is satisfied when $\phi$ lies in the range
$0 < \phi < \frac {\pi}{2}$ ($- \frac {\pi}{2} < \phi < 0$)
for the choice $\chi >0$ ($\chi < 0$). Thus any nonzero value of $\phi$
within its fundamental region can generate a $2$-soliton state. 
Looking at the momentum $P$ in (\ref{b19}), we see that the ratio $P/\phi >0$
since $\phi$ and $\chi$ have the same sign. Thus the chirality property
of the classical solitons is preserved in the quantum theory for $N=2$.
 
We will now consider the more interesting case with $N \geq 3$. 
Due to the parity symmetry of the Hamiltonian in (\ref {b3}),
we will henceforth assume that $\phi > 0$.
The inequalities in Eq. (\ref{b18}) can therefore be rewritten as
\beq
\chi ~\sin (l\phi) ~\sin [(N-l) \phi] ~>~ 0 \quad {\rm for} \quad
l ~=~ 1, ~2, ~\cdots ~, N-1 ~, 
\label{c1}
\eeq
or
\beq
\chi ~[~ \cos [ (N-2l) \phi ] ~-~ \cos (N\phi) ~] ~>~ 0 \quad {\rm for} \quad
l ~=~ 1, ~2, ~\cdots ~, N-1 ~. 
\label{c2}
\eeq
It is now convenient to consider the cases $\chi > 0$ and $\chi < 0$ 
separately.

For $\chi > 0$, Eq. (\ref{c2}) takes the form
\beq
\cos [ (N-2l) \phi ] ~>~ \cos (N\phi) \quad {\rm for} \quad
l ~=~ 1, ~2, ~\cdots ~, N-1 ~. 
\label{c3}
\eeq
Let us now consider a value of $\phi$ of the form
\beq
\phi_{N,n} ~\equiv ~\frac{\pi n}{N} ~,
\label{phinn}
\eeq
where $n$ is an integer satisfying $1 \le n < N/2$. Now, if $n$ is odd,
$\cos (N \phi_{N,n}) = -1$ and it is possible that all the inequalities in
(\ref{c3}) will be satisfied since $-1$ is the minimum possible value
of the cosine function. However, a closer look reveals that all the
inequalities in (\ref{c3}) are satisfied only if $N$ and $n$ are 
relatively prime, i.e., if the greatest common divisor of $N$ and $n$ is 1.
If $N$ and $n$ are not relatively prime, then let $p$ be an integer (greater
than 1) which divides both of them. Consider the integers $N' = N/p$ and
$n' = n/p$. Since $n$ is odd, $p$ must be odd and therefore $n'$ is also odd. 
Similarly, $N'=N/p$ is even (odd) if $N$ is even (odd), and therefore $N-N'$ 
is an even integer which is less than $N$. We then see that there is an 
integer $l$ for which $N-2l =N'$, and therefore $\cos [(N-2l) \phi_{N,n}] =
\cos (\pi n') =-1$; hence that particular inequality in Eq. (\ref{c3}) will 
be violated. We therefore conclude that all the inequalities in (\ref{c3}) 
will be satisfied for $\phi = \phi_{N,n}$ and $n$ odd, if and only if $N$ and 
$n$ are relatively prime. 

Similarly, we can consider the case $\chi < 0$. Eq. (\ref{c2}) then takes the 
form
\beq
\cos [ (N-2l) \phi ] ~<~ \cos (N\phi) \quad {\rm for} \quad
l ~=~ 1, ~2, ~\cdots ~, N-1 ~. 
\label{c5}
\eeq
It is clear that all of these cannot be satisfied if $N$ is even; in
particular the inequality with $l=N/2$ will fail. We therefore assume that
$N$ is odd. Once again, we consider a value of $\phi$ of the form given
in (\ref{phinn}) where $n$ is now even. Now $\cos (N \phi_{N,n}) = 1$ and 
there is a chance that all the inequalities in (\ref{c5}) will be satisfied 
since $1$ is the maximum possible value of the cosine function. However, we 
again find that this is true only if $N$ and $n$ are relatively prime.
If $N$ and $n$ are not relatively prime, then let $p$ be an integer (greater
than 1) which divides both of them. Consider the integers $N' = N/p$ and
$n' = n/p$. Since $N$ is odd, $p$ must be odd and therefore $N'$ is also odd;
hence $N-N'$ is an even integer which is less than $N$. Similarly, $n'=n/p$ 
is even since $n$ is even. Thus there is an integer $l$ for which $N-2l =N'$, 
and then $\cos [(N-2l) \phi_{N,n}] = \cos (\pi n') =1$; hence that particular 
inequality is violated in Eq. (\ref{c5}). We therefore conclude that all the 
inequalities in (\ref{c5}) will be satisfied for $\phi = \phi_{N,n}$ and $n$ 
even, if and only if $N$ and $n$ are relatively prime. 

Putting these statements together, we see that all the inequalities in 
(\ref{c2}) are satisfied for $\phi = \phi_{N,n}$, if and only if $N$ and $n$ 
are relatively prime (with $n$ odd for $\chi >0$, and $n$ even for $\chi < 0$).
By continuity, it then follows that all the inequalities will hold in a 
neighborhood of $\phi_{N,n}$ extending from a value $\phi_{N,n,-}$ to a
value $\phi_{N,n,+}$, such that $\phi_{N,n,-} < \phi_{N,n} < \phi_{N,n,+}$. We
will call the region
\beq
\phi_{N,n,-} < \phi < \phi_{N,n,+}
\eeq
as the band $B_{N,n}$. In this band, there is a soliton state with $N$ 
particles. 

For a given value of $N$, the number of bands in which soliton states exists
is equal to the number of integers $n$ which are relatively prime to $N$ and
satisfy $1 \le n < N/2$. This is equal to half the number of integers
which are relatively prime to $N$ and satisfy $1 \le n < N$. The latter
number is called Euler's $\phi$-function or totient function $\Phi (N)$
\cite{nive}. If $N$ has the prime factorization 
\beq
N ~=~ p_1^{n_1} ~p_2^{n_2} \cdots p_k^{n_k} ~,
\eeq
where $p_1, p_2, \cdots , p_N $ are all prime numbers,
then the totient function is given by
\beq
\Phi (N) ~=~ N \prod_{i=1}^k ~(~ 1 ~-~ \frac{1}{p_i} ~)~ .
\eeq
Thus the number of bands lying within the region $\eta>0$ coincides with
$\Phi (N) /2$. From the parity symmetry of Hamiltonian (\ref {b3}), 
it follows that the total number of bands lying within the full range
of $\eta$ is given by $\Phi(N)$. 

We now have to determine the end points $\phi_{N,n,-}$ and $\phi_{N,n,+}$
of the band $B_{N,n}$. The original inequalities in (\ref{c1}) show that
the end points are given by $\phi$ of the form 
\beq
\phi ~=~ \frac{\pi j}{l} ~,
\label{jl}
\eeq
where $j$ and $l$ are integers satisfying 
\beq
1 ~\le ~ l ~<~ N ~, \quad {\rm and} \quad j ~< ~\frac{l}{2}
\label{c6}
\eeq
(since $\phi < \pi /2$). In general, these $j$ and $l$ may not be relative 
prime numbers. However, one can always choose two relative primes $j'$ and 
$l'$ satisfying the constraint (\ref {c6}) such that $j'/l' = j/l$.
Thus the end points of the band $B_{N,n}$ are
given by two rational numbers $\phi /\pi$ of the form $j/l$ (where $j, ~l$
are relative primes satisfying the conditions in (\ref{c6})) which lie {\it 
closest} to (and on either side of) the point $\phi_{N,n} /\pi = n/N$. 
The solution to this problem is well known in
number theory and is described by the Farey sequences \cite{nive}.

For a positive integer $N$, the Farey sequence is defined to be the set of all
the fractions $a/b$ in increasing order such that (i) $0 \le a \le b \le N$,
and (ii) $a$ and $b$ are relatively prime. The Farey sequences for the first
few integers are given by
\bea
F_1: & & \quad \frac{0}{1} ~~~~\frac{1}{1} \non \\
F_2: & & \quad \frac{0}{1} ~~~~\frac{1}{2} ~~~~\frac{1}{1} \non \\
F_3: & & \quad \frac{0}{1} ~~~~\frac{1}{3} ~~~~\frac{1}{2} ~~~~\frac{2}{3} ~~~~
\frac{1}{1} \non \\
F_4: & & \quad \frac{0}{1} ~~~~\frac{1}{4} ~~~~\frac{1}{3} ~~~~\frac{1}{2} ~~~~
\frac{2}{3} ~~~~\frac{3}{4} ~~~~\frac{1}{1} \non \\
F_5: & & \quad \frac{0}{1} ~~~~\frac{1}{5} ~~~~\frac{1}{4} ~~~~\frac{1}{3} ~~~~
\frac{2}{5} ~~~~\frac{1}{2} ~~~~\frac{3}{5} ~~~~\frac{2}{3} ~~~~
\frac{3}{4} ~~~~\frac{4}{5} ~~~~\frac{1}{1}
\eea
These sequences enjoy several properties, of which we list the relevant ones
below.

\noindent (i) $a/b < a'/b'$ are two successive fractions in a Farey 
sequence $F_N$, if and only if
\beq
a' b ~-~ a b' ~=~ 1 ~, \quad {\rm and} \quad b ~,~ b' ~\le ~N ~.
\label{c7}
\eeq
It then follows that both $a$ and $b'$ are relatively prime to $a'$ and $b$.

\noindent (ii) For $N \ge 2$, if $n/N$ is a fraction appearing somewhere in the
sequence $F_N$ (this implies that $N$ and $n$ are relatively prime according to
the definition of $F_N$), then the fractions $a_1/b_1$ and $a_2/b_2$ appearing
immediately to the left and to the right respectively of $n/N$ satisfy
\bea
a_1 ~,~ a_2 ~\le ~n ~, \quad {\rm and} \quad a_1 ~+~ a_2 ~=~ n ~, \non \\
b_1 ~,~ b_2 ~<~ N ~, \quad {\rm and} \quad b_1 ~+~ b_2 ~=~ N ~.
\label{c8}
\eea

To return to our problem, we now see that the points $\phi_{N,n}$ in
(\ref{phinn}) (which
lie in the bands $B_{N,n}$) have a one-to-one correspondence with the 
fractions $n/N$, which appear on the left side of $1/2$ within the sequence 
$F_N$. Due to Eqs. (\ref{jl}) and (\ref {c6}), the end points of the band 
$B_{N,n}$ are given by
\bea
\phi_{N,n,-} ~=~ \frac{\pi a_1}{b_1} ~, \non \\
\phi_{N,n,+} ~=~ \frac{\pi a_2}{b_2} ~,
\label{c9}
\eea
where $a_1/b_1$ and $a_2/b_2$ are the fractions lying immediately to the left
and right of $n/N$ in the Farey sequence $F_N$. These
are the two unique fractions which lie closest to (and on either side) of
$n/N$ whose denominators are less than $N$ (due to (\ref{c8})). Therefore,
as we move away from $\phi = \phi_{N,n}$, one of the inequalities in 
(\ref{c1}) will be violated for the first time at these two points. (The
property $b_1 + b_2 = N$ shows that the same inequality in (\ref{c1}) is
violated at the two end points of a given band $B_{N,n}$).

The end points of a band given in Eq. (\ref{c9}) satisfy the property
\bea
n b_1 ~-~ N a_1 &=& 1 ~, \non \\
n b_2 ~-~ N a_2 &=& - ~1 ~,
\label{c10}
\eea
due to the property (\ref{c7}) of Farey sequences. Given two integers $N$
and $n$ satisfying the conditions given above, one can 
numerically find the integers $a_i$ and $b_i$ by searching for solutions
of Eqs. (\ref{c10}) within the limits given in (\ref{c8}). For the 
case of lowest band ($n=1$), these equations can also
be solved analytically to yield: $a_1=0,~b_1= 1,~ a_2=1,~ b_2= N-1$.
Thus, for every $N \ge 3$, the range of lowest band is given by $0 < 
\phi/\pi < 1/(N-1)$ which agrees with the result obtained in Ref. \cite{basu3}.
It turns out that there is a way to analytically determine the integers $a_i$ 
and $b_i$ for the case of a general $n$ by
using the idea of continued fractions; this will be described in Sec. 5.

In Table 1, we show the ranges of values of $\phi$ for which solitons exist 
for $N=2$ to 9. In Fig. 1, we present the same information pictorially for $N$
going up to 20.

\vspace{0.4cm}

\begin{center}
\begin{tabular}{|c|c|c|} \hline
$N$ & $n$ & Range of values \\
& & of $\phi /\pi$ \\ \hline
2 & 1 & $0 < \phi /\pi < 1/2$ \\
3 & 1 & $0 < \phi /\pi < 1/2$ \\
4 & 1 & $0 < \phi /\pi < 1/3$ \\
5 & 1 & $0 < \phi /\pi < 1/4$ \\
5 & 2 & $1/3 < \phi /\pi < 1/2$ \\
6 & 1 & $0 < \phi /\pi < 1/5$ \\
7 & 1 & $0 < \phi /\pi < 1/6$ \\
7 & 2 & $1/4 < \phi /\pi < 1/3$ \\
7 & 3 & $2/5 < \phi /\pi < 1/2$ \\
8 & 1 & $0 < \phi /\pi < 1/7$ \\
8 & 3 & $1/3 < \phi /\pi < 2/5$ \\
9 & 1 & $0 < \phi /\pi < 1/8$ \\
9 & 2 & $1/5 < \phi /\pi < 1/4$ \\
~~9~~ & ~~4~~ & ~~$3/7 < \phi /\pi < 1/2$~~ \\ \hline
\end{tabular}
\end{center}
\vspace{0.2cm}

\centerline{Table 1. The range of values of $\phi /\pi$ for which solitons
exist for various values of $N$.}
\vspace{0.4cm}

{} Due to the relations in (\ref{c10}), we see that the width of the
right side of the band $B_{N,n}$ from $\phi_{N,n}$ to $\phi_{N,n,+}$ is
$\pi /(N b_2)$, while the width of the left side from $\phi_{N,n,-}$ to 
$\phi_{N,n}$ is $\pi /(N b_1)$. For later use, we note that each of
these widths is strictly larger than $\pi /N^2$, since $b_1, ~b_2 < N$.
The total width $W_{N,n}$ of the band $B_{N,n}$ is given by 
\beq
W_{N,n} ~=~ \frac{\pi}{Nb_1} ~+~ \frac{\pi}{Nb_2} ~.
\label{wnn}
\eeq
For a given value of $N$, we will now find an expression for the total width 
of all the bands.

Consider two different bands $B_{N,n}$ and $B_{N,n'}$ and their end points 
given by 
\bea
\frac{a_1}{b_1} &<& \frac{n}{N} ~<~ \frac{a_2}{b_2} ~, \non \\
\frac{a'_1}{b'_1} &<& \frac{n'}{N} ~<~ \frac{a'_2}{b'_2} ~,
\eea
where we know from (\ref{c7}) that $n, ~n', ~b_1, ~b_2, ~b'_1, ~b'_2$ are all
relatively prime to $N$. We also know that $n, ~n' < N/2$; we will assume 
that $n'<n$. We will now show that $b'_1$ and $b'_2$ are not equal to 
either $b_1$ or $b_2$. Eq. (\ref{c7}) implies that
\bea
Na_1 ~-~ nb_1 &=& -1 ~, \non \\
Na_2 ~-~ nb_2 &=& 1 ~, \non \\
Na'_1 ~-~ n' b'_1 &=& -1 ~.
\eea
We then see that 
\bea
N ~(a_1 - a'_1) &=& nb_1 - n' b'_1 ~, \non \\
N ~(a_2 + a'_1) &=& nb_2 + n' b'_1 ~.
\eea
If $b'_1$ was equal to $b_1$, we would get $N(a_1 - a_1') = b'_1 (n-n')$, i.e.,
$N$ is a factor of $b'_1 (n-n')$.
Since $b'_1$ is relatively prime to $N$, this would mean that $N$ must be a 
factor of $n-n'$ \cite{nive}. But this is not possible since $n-n' < N$.
Similarly, if $b'_1$ was equal to $b_2$, we would get $N(a_2 + a_1') = b'_1
(n+n')$, i.e., $N$ 
is a factor of $n+n'$. But this is also not possible since $n+n' < N$.
We therefore conclude that $b'_1$ is not equal to $b_1$ or $b_2$. 
Similarly, we can show that $b'_2$ is not equal to $b_1$ or $b_2$.

We thus have the result that for any value of $N$, the denominators of the 
end points of the various bands $B_{N,n}$ are all different from each other;
we also know that they are all smaller than and relatively prime to $N$. 
Since the numbers of bands is equal to half the number of integers less than 
and relatively prime to $N$, and each band has two end points, the set of 
denominators of the end points of all the bands must be exactly the same as
the set of integers less than and relatively prime to $N$. Eq. (\ref{wnn}) now
implies that the total width of all the bands is given by
\beq
W_N ~=~ \frac{\pi}{N} ~\sum_l ~\frac{1}{l} ~,
\eeq
where the sum runs over all values of $l$ which are relatively prime to 
$N$ and satisfy $1 \le l \le N-1$. If $N$ is prime, we obtain 
\bea
W_N ~=~ \frac{\pi}{N} ~\sum_{l=1}^{N-1} ~\frac{1}{l} ~
& \simeq & \frac{\pi}{N} ~[~ \ln (N-1) ~+~ \gamma ~] \quad {\rm as} \quad
N \rightarrow \infty ~,
\eea
where $\gamma = 0.57721 \cdots$ is Euler's constant. 
We have numerically studied the behaviors of $W_N$ and the sum 
$I_N \equiv \sum_{n=2}^N W_N$ for values of $N$ up to $10^4$. We find
that although $W_N$ is a non-monotonic function of $N$, on the average
it grows with $N$ as $\ln N /N$; consequently, $I_N$ grows as
$(\ln N)^2$ as $N$ becomes large. [In the range $N=10^2$ to $10^4$, we find
that $NW_N/[\pi (\ln (N-1) + \gamma)]$ fluctuates between 0.255 and 1.001,
while $W_N /(\ln N)^2 = 1.179$ for $N=10^4$].

We thus see that the allowed range of values of $\phi$ for which solitons 
exist goes to zero generically as $\ln N /N$ as $N \rightarrow \infty$. In
contrast to this, the lowest band runs from $0$ to $\pi/(N-1)$ and
therefore has a width of $\pi/(N-1)$. This implies that the width of the
lowest band becomes an insignificant fraction of the total width as $N$
becomes large.

\vspace{1cm}

\noindent \section{Momentum and binding energy of a $N$-soliton state}
\renewcommand{\theequation}{4.{\arabic{equation}}}
\setcounter{equation}{0}

\medskip

In this section, we will calculate the momentum and binding energy for the 
$N$-soliton states described above. 
We first look at the momentum of the solitons in a particular band $B_{N,n}$
using Eq.(\ref{b19}). The form of the end points given in (\ref{c9}) shows 
that $\sin (N \phi) =0$ at only one point in the band $B_{N,n}$, namely, at 
$\phi = \phi_{N,n}$. In the right part of the band (i.e., from $\phi_{N,n}$
to $\phi_{N,n,+}$), the sign of $\sin (N\phi)$ is $(-1)^n$. In the left part 
of the band (i.e., from $\phi_{N,n,-} $ to $\phi_{N,n}$), the sign of $\sin 
(N\phi)$ is $(-1)^{n+1}$. Now, the analysis given above showed that $\chi$ 
has the same sign as $(-1)^{n+1}$. Hence the momentum given in (\ref{b19}) is 
positive in the left part of the band, negative in the right part of 
the band, and zero at $\phi = \phi_{N,n}$. 

Next, we look at the energy using Eq. (\ref{b20}). 
To calculate the binding energy, we consider a reference state in which the 
momentum $P$ of the $N$-soliton state given in (\ref{b19}) is equally 
distributed among the $N$ single-particle scattering states. The real 
wave number associated with each of these single-particle states is denoted 
by $k_0$. From Eqs. (2.10a) and (\ref{b19}), we obtain 
\beq
k_0 ~=~ \frac{\chi\sin(N \phi)}{N\sin \phi}. 
\label{d2}
\eeq
Using Eq. (2.10b), we can calculate the total energy for 
the $N$ single-particle scattering states as
\bea
E_s ~=~ \hbar^2 N k_0^2 ~=~ \frac{\hbar^2 \chi^2 \sin^2 (N \phi)}{N\sin^2 
\phi} ~.
\label{d3}
\eea
Subtracting $E$ in (\ref{b20}) from $E_s$ in (\ref{d3}), we obtain the 
binding energy of the $N$-soliton state as 
\beq
E_B (\phi, N) \equiv E_s - E ~=~ \frac{\hbar^2 \chi^2\sin (N \phi)}{\sin \phi}
\Big\{\frac{\sin (N \phi)}{N\sin \phi} -\frac{\cos (N \phi)}{\cos \phi} 
\Big\} ~.
\label{d4}
\eeq
It may be noted that the above expression of binding energy remains 
invariant under the transformation $\phi \rightarrow -\phi$. 

Substituting $N=2$ in Eq. (\ref{d4}), we obtain $E_B(\phi, 2) =2\hbar^2 \chi^2
\sin^2 \phi$. Thus we get $E_B (\phi,2) >0$ for any nonzero value of $\phi$. 
Let us now look at the case $N \ge 3$. We can rewrite Eq. (\ref{d4}) in the
form
\bea
E_B (\phi ,N) &=& \frac{\hbar^2 \chi \sin (N \phi)}{N \sin^2 \phi \cos \phi} ~
f (\phi , N) ~, \non \\
f (\phi , N) &=& \chi ~[~ \sin (N \phi) \cos \phi ~-~ N \cos (N \phi) \sin 
\phi ~]~ .
\label{d5}
\eea
We will now prove that the function $f (\phi , N)$ is positive in all the
bands $B_{N,n}$ for all values of $N$ and $n$. To show this, we add up
all the inequalities given in (\ref{c2}), and use the identities 
\bea
\sum_{l=1}^{N-1} ~\cos [(N-2l) \phi] &=& \frac{\sin [(N-1) \phi]}{\sin \phi} ~,
\non \\
\sin [(N-1) \phi] &=& \sin (N \phi) \cos \phi - \cos (N \phi) \sin \phi ~. 
\eea
We then find that $f (\phi ,N) > 0$. Hence, $E_B$
given in (\ref{d5}) has the same sign as $\chi \sin (N \phi)$. Following
arguments similar to that of the momentum, we find that the binding
energy is positive in the left part of each band (called bound states), 
negative in the right part (called anti-bound states), and zero at the 
point $\phi = \phi_{N,n}$.

We thus see that for $\phi >0$, the momentum and the binding energy are
both positive in the left part of each band, and they are both negative
in the right part. This is to be contrasted with the classical solitons
in the classical DNLS model which always have positive momentum. If
$\phi < 0$, we can similarly show that solitons with positive values of
$P/\phi$ have positive binding energy, and solitons with negative values
of $P/\phi$ have negative binding energy.

In Fig. 2, we show the binding energy $E_B$ in (\ref{d4}) as a function of 
$\phi /\pi$ for three different values of $N$. (We have set $\hbar^2 \chi^2 
=1$ in that figure). We see that $E_B$ is indeed positive (negative) in the 
left (right) part of each band.

\vspace{1cm}

\noindent \section{Continued fractions}
\renewcommand{\theequation}{5.{\arabic{equation}}}
\setcounter{equation}{0}

\medskip

In this section, we will apply the technique of continued fractions to our
problem. This will lead us to an explicit expression for the end points of
the band $B_{N,n}$ for any value of $N$ and $n$. We will also address the 
problem of determining the values of $N$
for which $N$-soliton states exist for a given value of $\phi$. It will turn
out that continued fractions provide a way of finding some (but not
necessarily all) values of $N$ for which solitons exist for a given $\phi$. 

We first discuss the idea of a continued fraction \cite{nive}. Any positive 
real number $x$ has a simple continued fraction expansion of the form
\beq
x ~=~ n_0 ~+~ \frac{1}{n_1 ~+~ \frac{1}{n_2 ~+~ \frac{1}{n_3 ~+~ \cdots}}} ~,
\eeq
where the $n_i$'s are integers satisfying $n_0 \ge 0$, and $n_i \ge 1$ for $i
\ge 1$. The expansion ends at a finite stage with a last integer $n_k$ if $x$ 
is rational. In that case, we can assume that the last integer satisfies $n_k 
\ge 2$. (If $n_k$ is equal to $1$, we can stop at the previous stage and 
increase $n_{k-1}$ by 1). With this convention for $n_k$, the continued 
fraction expansion is unique for any rational number $x$. If $x$ is an 
irrational number, the continued fraction expansion does not end, and it is 
unique. 

We will be concerned below with the continued fraction expansion for $x=\phi /
\pi$ which lies between 0 and 1/2; hence, $n_0 =0$. We will
use the notation $x=[0,n_1,n_2,n_3,\cdots]$ for such an expansion. 

Given a number $x$, the integers $n_i$ can be found as follows. We define
$x_0 = x$. Then $n_0 = [x_0]$, where $[y]$ denotes the integer part of a
non-negative number $y$. We then recursively define $x_{i+1} = 1/(x_i - n_i)$,
and obtain $n_{i+1} = [x_{i+1}]$ for $i =0,1,2,\cdots$. If we stop at the 
$k^{\rm th}$ stage, we obtain a rational number $r_k =[n_0,n_1,n_2,\cdots ,
n_k]$ which is an approximation to the number $x$. If we write $r_k = p_k/q_k$,
where $p_k$ and $q_k$ are relatively prime, then we have the following three 
properties for all values of $k \ge 1$ \cite{nive},
\bea
q_k &<& q_{k+1} ~, \non \\ 
p_{k+1} q_k ~-~ p_k q_{k+1} &=& (-1)^k ~, \non \\
|~ x ~-~ \frac{p_k}{q_k} ~| &<& \frac{1}{q_k^2} ~. 
\label{e1}
\eea
Using these properties, we can now find the end points $\phi_{N,n,\pm}$
of the band $B_{N,n}$. In Sec. 3, we saw that this is equivalent to 
finding the fractions $a_1/b_1$ and $a_2/b_2$ which lie to the immediate
left and right of the fraction $n/N$ in the Farey sequence $F_N$. Let
us suppose that $n/N$ has a continued fraction expansion given by
\bea
\frac{n}{N} &=& [0,n_1,n_2,\cdots,n_{k-1},n_k] \non \\
&=& [0,n_1,n_2,\cdots,n_{k-1},n_k -1,1] ~,
\label{e2}
\eea
where the expression in the second equation can be seen to be equivalent to 
the one in the first equation. Now consider the two fractions
\bea
\frac{a}{b} &=& [0,n_1,n_2,\cdots ,n_{k-1}] ~, \non \\
\frac{c}{d} &=& [0,n_1,n_2,\cdots ,n_{k-1},n_k -1] ~.
\label{e3}
\eea
On comparing these expressions to those given in (\ref{e2}) and using the
properties of continued fractions in (\ref{e1}), we see that
\bea
b, ~d &<& N ~, \non \\
nb ~-~ Na &=& (-1)^{k-1} ~, \non \\
nd ~-~ Nc &=& (-1)^k ~.
\eea
We now see from Eq. (\ref{c7}) that if $k$ is odd, then $a/b$ and $c/d$
are to the immediate left and right respectively of $n/N$ in the Farey
sequence $F_N$, while if $k$ is even, then $a/b$ and $c/d$
are to the immediate right and left respectively of $n/N$ in the sequence 
$F_N$. Hence the end points $\phi_{N,n,-} /\pi$ and $\phi_{N,n,+} /\pi$ 
are given by $a/b$ and $c/d$ respectively (in Eq. (\ref{e3})) if $k$ is odd,
and vice versa if $k$ is even. Thus the continued fraction expansion gives
a convenient way of determining $\phi_{N,n,\pm}$ if $N$ and $n$ are given.

As an explicit example, consider the value $\phi /\pi =3/8$. The continued
fractions expansion of this is given by $\phi /\pi = [0,2,1,2]=[0,2,1,1,1]$. 
Eq. (\ref{e3}) then gives the two neighboring fractions in the Farey sequence 
$F_8$ to be $a/b =[0,2,1]=1/3$ and $c/d=[0,2,1,1]=2/5$. The band $B_{8,3}$
therefore lies in the range $1/3 < \phi /\pi < 2/5$ as shown in Table 1.

Next, we study the question of determining the values of $N$ for
which $N$-body solitons exist if a value of $\phi$ is given. Suppose that 
we know the continued fraction expansion 
\beq
\frac{\phi}{\pi} ~=~ [0,n_1,n_2,\cdots ] ~.
\label{e4}
\eeq
Suppose that we stop at any any point in this expansion, say, at the 
$k^{\rm th}$ stage, and we get
\beq
\frac{p_k}{q_k} ~=~ [0,n_1,n_2,\cdots , n_k] ~.
\eeq
Then we know from (\ref{e1}) that 
\beq
|~ \frac{\phi}{\pi} ~-~ \frac{p_k}{q_k} ~| ~<~ \frac{1}{q_k^2} ~.
\label{e5}
\eeq
We now recall from Sec. 3 that both the right and the left part of
the band $B_{q_k,p_k}$ have widths which are larger than $1/q_k^2$.
Eq. (\ref{e5}) therefore implies that $\phi /\pi$ must lie within the 
band $B_{q_k,p_k}$. We have thus found a value of $N=q_k$ for which we
know that a $N$-body soliton state exists for the given value of $\phi$.
We can generate several such values of $N$ by stopping at different stages
$k$ in the expansion given in (\ref{e4}). 

If $\phi /\pi$ is rational, the
continued fraction expansion stops at a finite stage, so we only obtain
a finite number of values of $N$ in this way. This can also be seen directly 
from Eq. (\ref{c1}). If $\phi /\pi = p/q$ is rational, then at least one of 
the inequalities in (\ref{c1}) will be violated if $N > q$. Moreover, one 
gets $k_n= k_{n+q}$ from Eq.(\ref {b16}) by putting $\phi /\pi = p/q$.
Thus at least two wave numbers would coincide when $N>q$. As a result, the 
eigenfunction (\ref {b8}) becomes trivial in this case.
We thus conclude that if $\phi /\pi$ is rational, there is only a finite
number of values of $N$ for which a $N$-body soliton state exists. On the
other hand, if $\phi /\pi$ is irrational, then the expansion in (\ref{e4})
does not end, and we can use the procedure described above to find an 
infinite number of possible values of $N$ for which a $N$-body soliton exists.

Given a value of $\phi$, the procedure described above will find values of $N$
for which we can prove that a $N$-body soliton exists. This does not rule
out the possibility that there may be other values of $N$ for which such
solitons exits. [To give a simple example, consider the value $\phi =\pi /10$.
Using Eq. (\ref{c1}), we see that there are bound states for all values of
$N$ from 2 to 10. However, the continued fraction technique only gives the
value of $N=10$ for which a soliton exists]. The continued fraction technique
therefore provides a sufficient but not necessary condition for finding the 
desired values of $N$.

\vspace{1cm}

\noindent \section{Conclusion}

\medskip

By applying the coordinate Bethe ansatz, we have investigated
the range of the coupling constant $\eta$ for which localized
quantum $N$-soliton states exist in the DNLS model. Using the ideas of
Farey sequences and continued fractions, we have given explicit expressions
for all the allowed ranges (bands) of $\eta$ for which $N$-body solitons
exist. The continued fraction also provides a way of finding some values of
$N$ for which soliton states exist for a given value of $\eta$. If
$\phi =\tan^{-1} (\eta)$ is rational (irrational), the number of values of $N$ 
for which soliton states exist is finite (infinite).

We find that for $N \ge 3$, the $N$-body solitons can have both positive and 
negative momentum. Thus the chirality property of classical DNLS solitons is 
generally not preserved at the quantum level. We also calculated the binding 
energy for the soliton states. We find that solitons with positive values of 
$P/\eta$ form bound states with positive binding energy. Solitons with 
negative values of $P/\eta$ have negative binding energy and hence form 
anti-bound states. The anti-bound solitons would be expected to be 
unstable in the presence of external perturbations. As a 
future study, it may be interesting to calculate the decay rate of such 
quantum solitons by introducing small perturbations in the DNLS Hamiltonian.

In the continuum version of the DNLS model (this can be obtained from the
$N$-body Schr\"odinger model by taking the limit $N \rightarrow \infty$), it 
is known that solitons exist only if $0 < N \v \phi \v < \pi$ 
\cite{shni,min}; this corresponds to the
lowest band. One might ask whether there is a continuum 
version of the solitons in the higher bands which we have found in this paper.
This may be an interesting problem for future study. It is possible that
only the lowest band has a counterpart in the continuum theory. Note that if 
$\phi$ lies in the lowest band, then the phases of $k_n$ in Eq. (\ref{b16})
vary slowly with $n$; the spacing between two neighboring phases is of order 
$1/N$. This may be necessary to ensure that the continuum limit exists. For 
the generic band, where $\phi$ is of order 1 rather than of 
order $1/N$, the phases of neighboring values of $k_n$ vary rapidly; this 
may make it difficult to take the continuum limit.

To conclude, we have shown that
the soliton structure in the DNLS model is much richer than in
the well known NLS model. In the latter model, the interaction part of the
Hamiltonian is given by $\mu \sum_{l<m} \delta (x_l - x_m)$ in the
coordinate representation. This model has $N$-soliton states for any
value of $N \ge 2$ and any value of $\mu < 0$; all the solitons have
positive binding energy. In contrast to this, the DNLS model allows soliton
states in only certain bands of the coupling constant for $N \ge 3$, and
these bands form intricate number theoretic patterns.

\newpage

\begin{figure}[htb]
\begin{center}
\epsfig{figure=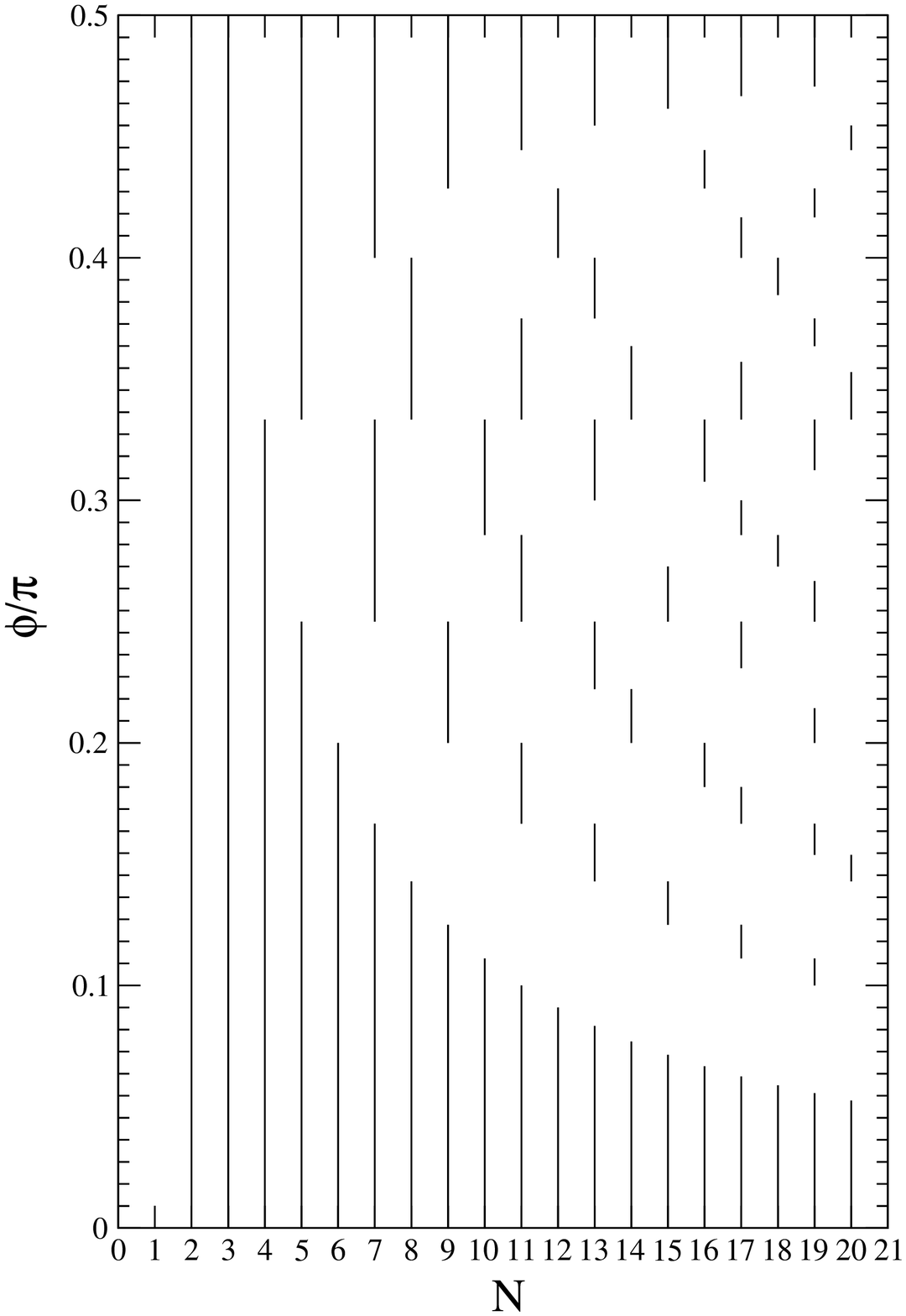,height=19cm,width=15cm}
\end{center}
\caption{The values of $\phi /\pi$ for which $N$-body soliton states exist 
for various values of $N$.}
\label{fig1}
\end{figure}

\newpage

\begin{figure}[htb]
\begin{center}
\epsfig{figure=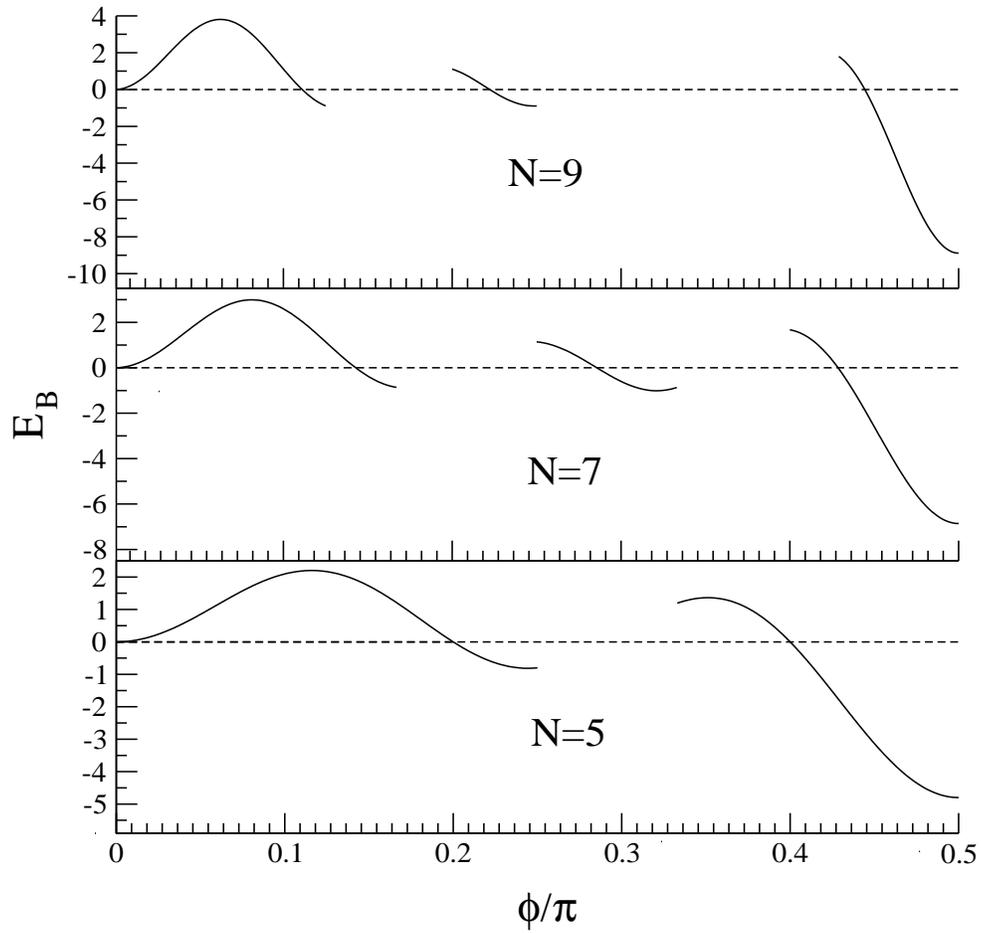,height=19cm,width=14cm}
\end{center}
\caption{The binding energy $E_B$ of the $N$-body soliton states as a function
of $\phi /\pi$ for three different values of $N$. Positive and negative values
of $E_B$ denote bound and anti-bound states respectively.}
\label{fig2}
\end{figure}

\end{document}